\documentclass[twocolumn,showpacs,preprintnumbers,amsmath,amssymb]{revtex4}

\usepackage{graphicx}
\usepackage{dcolumn}
\usepackage{bm}

\begin{document}


\title{Interplay between Fe 3\textit{d} and Ce 4\textit{f} magnetism and Kondo interaction in CeFeAs$_{1-x}$P$_{x}$O probed
\\by $^{75}$As and $^{31}$P NMR}
\author{R Sarkar}
\altaffiliation[]{rajib.sarkar@cpfs.mpg.de, rajibsarkarsinp@gmail.com}
\author{M Baenitz}
\author{A Jesche}
\author{C Geibel}
\author{F Steglich}
\affiliation{
Max-Planck Institute for Chemical Physics of Solids, 01187
Dresden, Germany
}%


\date{\today}

\begin{abstract}
A detailed $^{31}$P (I=1/2) and $^{75}$As (I=3/2) NMR study on
polycrystalline CeFeAs$_{1-x}$P$_{x}$O alloys is presented. The
magnetism of CeFeAsO changes drastically upon P substitution on
the As site.  CeFePO is a heavy fermion system without long range
order whereas CeFeAsO exhibits a Fe-3$d$ SDW type of ordering
accompanied by a structural transition from tetragonal ($TT$) to
orthorhombic ($OT$) structure. Furthermore Ce-4$f^{1}$ orders
antiferromagnetically (AFM) at low temperature. At the critical
concentration where the Fe-magnetism is diminished the Ce-Ce
interaction changes to a ferromagnetic (FM) type of ordering.
Three representative samples of the CeFeAs$_{1-x}$P$_{x}$O
($x$=0.05, 0.3 and  0.9) series are systematically investigated.
1) For the $x$=0.05 alloy a drastic change of the line width at
130 K indicates the AFM-SDW type of ordering of Fe and the
structural change from $TT$ to $OT$ phase. The line width roughly
measures the internal field in the ordered state and the
transition is most likely first order. The small and nearly
constant shift frm $^{31}$P and $^{75}$As NMR suggests the
presence of competing hyperfine interactions between the nuclear
spins and the 4$f$ and 3$d$ ions of Ce and Fe.  2) For the
$x$=0.3 alloy, the evolution of the Fe-SDW type of order takes
place at around 70 K corroborating the results of bulk
measurement and $\mu SR$. Here we found evidence for phase
separation of paramagnetic and magnetic  SDW phases. 3) In
contrast to the heavy fermion CeFePO for the $x$=0.9 alloy a phase
transition is found at 2 K. The field-dependent NMR shift gives
evidence of FM ordering. Above the ordering the spin lattice
relaxation rate $^{31}$(1/T$_{1}$) shows unconventional,
non-Korringa like behaviour which indicates a complex interplay
of Kondo and FM fluctuations.

\end{abstract}

\maketitle
\section{\label{sec:level1}Introduction:}
\begin{figure}
\includegraphics[scale=.40]{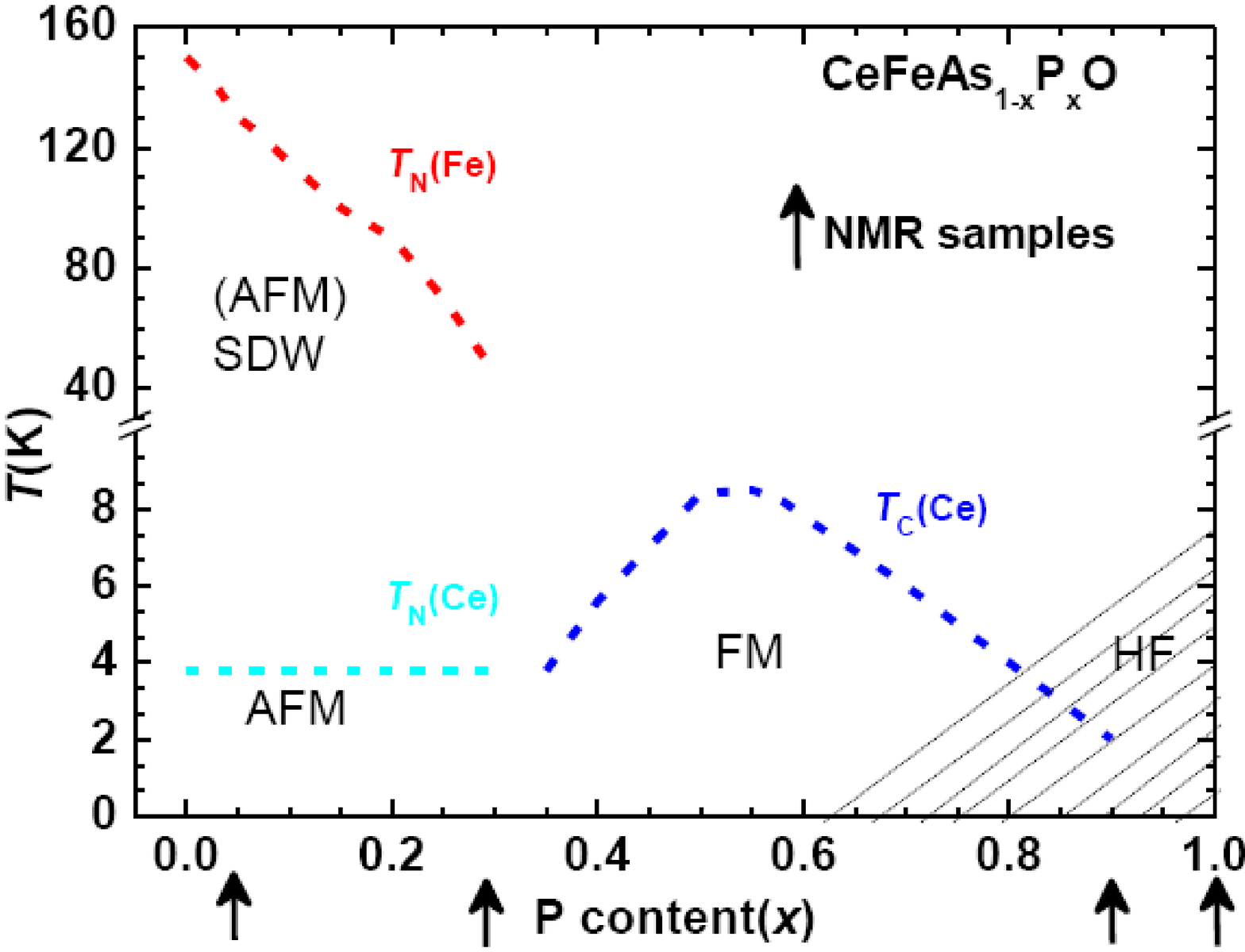}
\caption{\label{fig:phasediagram} Schematic phase diagram based
on the Refs. \cite{{Anton-CeFeAsPO-to be published-2009},{Yongkang
Luo-condmat-2009},{Clarina de la
Cruz-condmat-CeFeAs1-xPxO-2009},{Jesche-PHD-Thesis-2011}}. NMR
samples are marked by $\uparrow$.}
\end{figure}

The rare earth transition metal pnictides $RTPn$O ($R$: rare
earth, $T$: transition metal, $Pn$: P or As) attracted
considerable attention because of the startling discovery of
superconductivity (SC) in the $R$FeAsO$_{1-x}$F$_{x}$ series of
compounds at elevated temperatures above $50~$K that are
surpassed only in the cuprate superconductors
{\cite{{Kamihara-2008},{Chen-2008},{Jun Zhao-nature
materials-2008},{Z. A. Ren-2008},{Chen-Li-2008},{J.
Yang-Supercond.Sci. Technol.-2008},{J. G. Bos- Chem.
Commun.-2008}}}.  In CeFeAsO, antiferromagnetic (AFM) ordering is
achieved by the Ce moments at $3.7~$K whereas the
high-temperature region is dominated by the $3d$ magnetism of Fe
which culminates in a SDW type of AFM order of Fe at $\sim\!$145
K. Interestingly, there is a structural transition from a
tetragonal ($TT$) to an orthorhombic ($OT$) phase at $\sim$150 K.
Neutron scattering and muon spin relaxation $\mu SR$ experiments
suggest that there is a sizeable inter-layer coupling in CeFeAsO
\cite{{Anton-CeFeAsO-NJP-2009},{Chi S-prl-2008},{Maeter
condmat-2009}}. Recent studies of CeRuPO and CeOsPO
\cite{{Krellner-2007},{Krellner-Crystal frowth-2007},{C.
Krellner-prl-2007}}, on the other hand, indicate that CeRuPO is a
rare example of a ferromagnetic (FM) Kondo lattice showing FM
order of Ce at $T_{\mathrm{C}}$=15 K and a Kondo energy scale of
about $T_\mathrm{K}$ $\simeq$ 10 K. In contrast, CeOsPO shows AFM
order at $T_{\mathrm{N}}$=4.5 K. However, recent studies of the Fe
derivative CeFePO suggest that this is a heavy fermion metal
close to a FM instability in which the magnetism is dominated by
the $4f$ electrons \cite{E. M. Bruning-prl-2008}.
\\
\indent The Fe $3d$ magnetism in these systems is very sensitive to the electronic environment.
The interplay between SC and magnetism was vividly demonstrated by the electron doping studies
in CeFeAsO$_{1-x}$F$_{x}$ ($T_{c} \simeq 41 K$ \cite{Shen V. Chong-2009}) as well as in pressure
 studies on optimally doped CeFeAsO\cite{Takeshitai-JPSJ-2009}. In this work we broaden investigations
 into electronic phenomena and the participation of electrons superconductivity and magnetism in the
  CeFeAsO system by substituting As in CeFeAsO by the smaller P ion. This approach seems especially
attractive as it involves strong electronic correlations in
heavy-fermion CeFePO. Theoretical work suggests that the Fe
pnictide series where P is progressively substituted for As may
present a route to magnetic quantum criticality \cite{Jianhui
Dai-nature mater-2008}. In this context, the substitutional
series CeFeAs$_{1-x}$P$_{x}$O appears to be especially attractive
presenting a crossover from AFM fluctuations to FM fluctuations.
This is already evidenced by the recent research work done by
Luo. et. al \cite{Yongkang Luo-condmat-2009}. Additional work
\cite{Anton-CeFeAsPO-to be published-2009}, from which the phase
diagram shown in Fig. ~\ref{fig:phasediagram} has been reproduced,
illustrates the appealing standoff between cooperative (SDW, AFM,
FM) and correlational (Kondo) phenomena in doped
CeFeAs$_{1-x}$P$_{x}$O, which is the topic of this work. The
primary macroscopic investigations of these alloys suggest that
by increasing the P doping concentration the Fe-SDW type anomaly
shifts to lower temperature and is completely suppressed at the
critical concentration of $x_{\mathrm{c}}$=0.35\cite{Clarina de la
Cruz-condmat-CeFeAs1-xPxO-2009}. Here, in addition, the nature of
the Ce $ 4f$ magnetism changed from AFM to FM. Between $x$=0.35 to
$x$=0.9 the ground state is governed by the long range Ce $4f$
based FM. Fe magnetism is absent beyond $x$=0.35. Above about
$x\approx$0.9 long range FM Ce order disappears and makes way to
the Kondo interaction. This is in line with our earlier
findings\cite{E. M. Bruning-prl-2008} which conveyed the
heavy-fermion nature close to FM order in the $x$=1 system,
CeFePO. In fact, the system appears to be partial to FM order as
is demonstrated by appropriately small elemental substitution (As
on the P site or Ru on the Fe site)\cite{Krellner-2007}. In this
paper we report $^{31}$P (I=1/2) and $^{75}$As (I=3/2) NMR
studies on CeFeAs$_{1-x}$P$_{x}$O with $x$= 0.05, 0.3, and 0.9
(see Fig. ~\ref{fig:phasediagram}) as a function of field and
temperature. Whereas there is already an abundance of research
reported on NMR work on $R$Fe$Pn$O type Fe pnictides involving
various elements in the place of $R$, to our knowledge there
exists at present only one paper concerning $^{75}$As$-$NMR
studies specifically on the CeFeAsO compound. Morover, the
present work involving $^{31}$P$-$NMR provides an important
advance on existing NMR studies by conspicuously addressing the
role of the P-dopant in CeFeAs$_{1-x}$P$_x$O. The rather complex
$I=3/2$ spectra did not enable clarification of the Fe AFM (SDW
type) ordered state \cite{Ghoshray-2009}, -a topic which we
sought to address in this work and hence this work involving
$^{31}$P$-$NMR provides an important advance on existing NMR
studies. A very small P substitution ($x=0.05$) in CeFeAsO
affords the opportunity to probe $^{31}$P as a favorable $I=1/2$
nucleus with less complex spectra than those obtained in
$^{75}$As$-$NMR, yet without impacting severely on the magnetism
in CeFeAsO. Besides, employing an $I=1/2$ nucleus obviates the
need to account for quadrupolar interactions.

\section{Experimental}
Polycrystalline  CeFeAs$_{1-x}$P$_{x}$O samples were synthesized
using a Sn flux method in evacuated quartz tubes as described in
Ref \cite{{Anton-CeFeAsO-NJP-2009},{Anton-CeFeAsPO-to be
published-2009}} (where $x$ denotes the nominal phosphorous
content). The phosphorus concentration was confirmed by EDX
analysis. X-ray powder diffraction  showed only tiny foreign
phases and the determined lattice spacings allowed for an
estimation of the P concentration by using Vegard's rule. For NMR
measurements, the powder samples were fixed in paraffin to ensure
a crystallographically random orientation and to prevent signal
reduction due to the skin-depth effect. $^{31}$P and $^{75}$As
NMR measurements were performed with a standard pulsed NMR
spectrometer (Tecmag) at the frequency 75 MHz ($^{31}$P-NMR) and
48 MHz ($^{75}$As-NMR) as a function of temperature. The
field-sweep NMR spectra were obtained by integrating the echo in
the time domain and plotting the resulting intensity as a
function of the field. Shift values are calculated from the
resonance field $H^{\ast}$ by $K(T)=(H_{L}-H^{\ast})/H^{\ast}$
whereas the Larmor field $H_{L}$ is given by using
H$_{3}$PO$_{3}$ ($^{31}$P-NMR) and GaAs ($^{75}$As-NMR) as
reference compounds. Spin-lattice relaxation time ($T_{1}$)
measurements were carried out by the standard saturation recovery
method. On the As rich side of the phase diagram the very short
$T_{1}$ lifetimes turned out too short to be measured, but
 no such difficulty was encountered at the
P-rich side ($x=0.9$). We profited from this fact by additionally
conducting field-dependent $^{31}$P-NMR in the $x=0.9$ sample in
order to probe the low-temperature Ce-based FM order.
\section{Results}
In this paper NMR results on the $^{31}$P and $^{75}$As nuclei on
three different CeFeAs$_{1-x}$P$_{x}$O samples are presented. The
first one ($x=0.05$) is close to the undoped end-point compound CeFeAsO. The midpoint concentration
 ($x=0.3$) is located at the border where Fe magnetism is suppressed.
The P-rich composition ($x=0.9$) was chosen close to the
heavy-fermion metal CeFePO as well as to enable detection of the
FM order. Therefore, the three samples represent appropriate
regions of interest on the complex phase diagram of this alloy
system. In the following section on $^{31}$P and $^{75}$As NMR
results are discussed in the context of magnetic and structural
transitions in the As-rich sample in particular, and the Kondo
interaction competing with the RKKY interaction in the As-poor
samples.
\subsection{\label{sec:level2}$^{31}$P and $^{75}$As NMR for \textit{x}=0.05: the SDW transition region }
\begin{figure}
\includegraphics[scale=1.05]{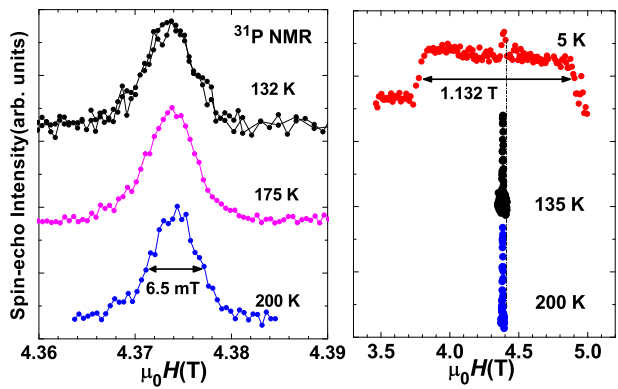}
\caption{\label{fig:PAs95P05} $^{31}$P field sweep NMR spectra
for $x=0.05$ at a selection of temperatures above the SDW transition
(left) and well inside the ordered region at $5~$K (right, expanded x-axis).}
\end{figure}

Figure~\ref{fig:PAs95P05} shows the $^{31}$P field sweep NMR
spectra at different temperatures. At high temperature a single
narrow line is observed, as expected for a TT system. On lowering
the temperature the overall features of the spectra remain
largely invariant down to $132~$K. No significant shift is
observed down to this temperature. With further lowering of
temperature to $130~$K, a weak signal-to-noise ratio develops
because of enormous line broadening in the $^{31}$P NMR spectra.
As a result we are unable to resolve $^{31}$P NMR spectra. This
is ascribed to SDW-type AFM ordering that develops abruptly
between $132$ and $130~$K, resulting in a reduction in $T_{1}$.
Nevertheless, towards lower temperatures at around $5~$K the
$^{31}$P NMR line re-emerged as a half-square like line shape.
\\
\indent The Full Width at Half Maximum (FWHM) of powder spectra in
the AFM ordered state provides an rough estimate of the
prevailing internal field of the system as sensed by the nuclei
being probed \cite{{Hideto-JPSJ-2008},{Baek-prb-2009}, {S.-H.
Baek-PRL-2009}}.
\\
\indent The horizontal double arrows in Fig.~\ref{fig:PAs95P05}
show the internal field at FWHM. Approaching the ordered phase
from the paramagnetic state evidently produces a rather drastic
change of the internal field. Although the overall spectra are
hardly shifted with respect to field upon lowering the
temperature. This is in contrast to $^{31}$P NMR finding for the
pure CeFePO system \cite{E. M. Bruning-prl-2008}. The
temperature-independent shift found for the sample $x=0.05$
indicates a cancelling out of hyperfine field contributions
originating from Ce and Fe at the P-site. Furthermore, the
cancellation could also be come by between the moments of the
Fe-sublattice themselves with P either occupying symmetric
positions in the magnetic lattice or random positions in an
incommensurate SDW.  The vertical line in the right-hand side
panel of Fig.~\ref{fig:PAs95P05} shows the position of the Larmor
field.
\begin{figure}
\includegraphics[scale=1.05]{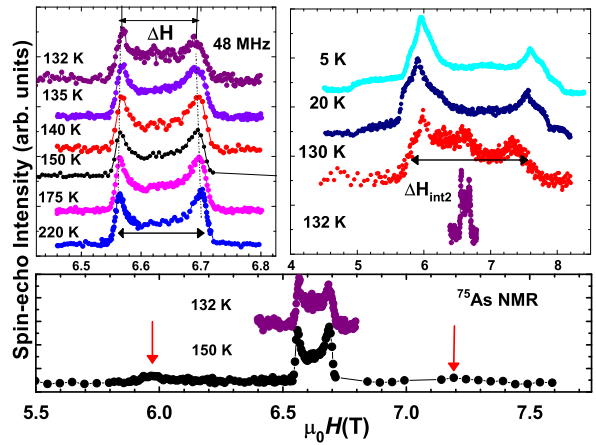}
\caption{\label{fig:AsAs_95P_05} $^{75}$As field sweep NMR
spectra at different temperatures for $x$=0.05. The upper left
panel shows the central transition with pronounced second-order
quadrupolar splitted $^{75}$As NMR spectra above the SDW
transition. Lower panel shows the $^{75}$As spectra at $150~$K
and $132~$K. Upper right panel shows the $^{75}$As spectra close
to ($132~$ and $130~$K) and below ($20~$ and $5~$K) the SDW
transition.}
\end{figure}
\\
\indent Figure ~\ref{fig:AsAs_95P_05} shows the $^{75}$As field
sweep NMR spectra at different temperatures (top part of the left
panel). The central transition of the $^{75}$As NMR spectra shows
a pronounced second-order quadrupolar splitting indicating
relatively large quadrupole interaction. The full $^{75}$As NMR
spectrum at $150~$K as well as the central part of the $132~$K
spectrum are shown at the lower panel. Arrows indicate the
location of satellite transitions. The upper right-hand panel
provides a comparison of the spectra above and below the SDW
transition. Typical powder response with strong quadrupole
coupling is evident. The NMR response evolves insipidly with
temperature down to $132~$K. Further cooling, however, results in
strong broadening of the resonant line and consequently
considerable losses in the signal-to-noise ratio. Similar
$^{75}$As NMR spectra were also reported for
NdFeAsO\cite{Jegli-PRB-2009} in which the line profiles were
explained in terms of the asymmetry parameter $\eta$ (here $\eta$=
(V$_{xx}$-V$_{yy}$)/V$_{zz}$, and V$_{xx}$, V$_{yy}$ and V$_{zz}$
represent the electric field gradients for three principle axes).
Interestingly, except for the weak resonance near $6.6~$T at
$T=130~$K, the nature of the line shape remains invariant down to
low temperatures.
\\
\indent The resonance at $6.6~$T, originating at high
temperatures and thus clearly a feature of the paramagnetic
phase, can be found down to $130~$K. Below this temperature the
line broadening signals SDW type of ordering, and one concludes
that in a narrow temperature window below SDW ordering the
paramagnetic phase and SDW-ordered phase may co-exist.
\begin{figure}
\includegraphics[scale=1.0]{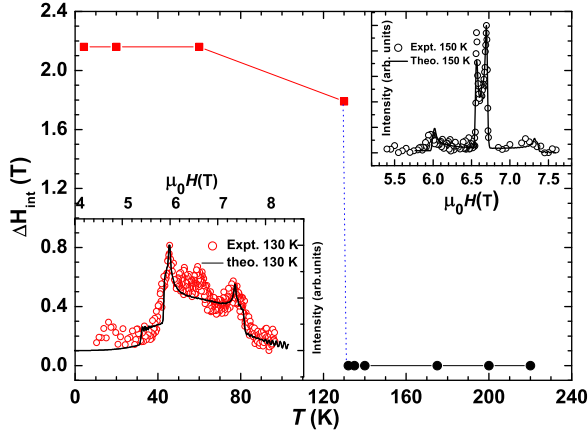}
\caption{\label{fig:Fit-Field-As_95P_05} Temperature dependence of
the FWHM of the $^{75}$As NMR spectra below (red square) and
above (closed circle) the SDW transition temperature. Right-hand
inset shows the $150~$K $^{75}$As spectra together with the
theoretical simulation. Left-hand inset shows the $130~$K
$^{75}$As spectra together with the theoretical simulation.}
\end{figure}

\indent The splitting $\Delta H_{Q}$ of the central NMR resonance
due to second-order quadrupolar interaction is given by
\cite{{George H. Stauss-JCP-1964},{Kenichiro
Tatsumi-JPSJ78-2009}};
\begin{equation}
\Delta H_{Q} =
\frac{\nu_{Q}^{2}}{48\gamma_{N}^{2}H}(25-22\eta+\eta^{2}),
\label{eq:nuQ}
\end{equation}
where $\gamma_{N}$ and $\nu_Q$ represents the gyromagnetic ratio
of $^{75}$As ($7.292~$MHz/$10~$kOe) and nuclear quadrupolar
splitting frequency, respectively. $\Delta H_{Q}$ depends on
$\nu_Q^{2}$, $1/H$ and $\eta$. $\Delta H _Q$ amounts to be
$0.132~$T above the SDW transition. This is shown in the
Fig.~\ref{fig:AsAs_95P_05} (top part of the left-hand panel).
\\
\indent  The insets of Fig.~\ref{fig:Fit-Field-As_95P_05} show
$^{75}$As NMR spectra at $150~$K and at $130~$K together with the
theoretical simulation. Simulation has been done considering the
standard 2nd order perturbation effect of quadrupolar interaction
incorporating the Knight-shift powder pattern. Furthermore we
have introduced Gussian line broadening effects
\cite{{G.C.Carter-Bennett},{Kitagawa-BaFe2As2-NMR77}}. At $150~$K
and $130~$K the spectra are fitted reasonably well, taking into
account the second-order quadrupolar perturbation contribution.
The parameters obtained for $150~$K are $\nu_{Q}=9.57 \pm
0.2$~MHz and $\eta=0.03$. These values are close to findings for
other members of the paramagnetic pnictide family
\cite{{Ghoshray-2009},{Mukuda-JPSP-2008},{Jegli-PRB-2009}}. For
$130~$K the parameters obtained are different, viz. $\nu_{Q}=11.00
\pm 0.4 $ MHz and $\eta=0.1 \pm 0.02$. This change reflects the
change of magnitude and direction of the electric field gradient
due to the structural phase transition. Using Eqn.\ \ref{eq:nuQ},
$\nu_{Q}=9.56$~MHz is obtained, which is very close to the value
obtained from a simulation of the entire $^{75}$As spectrum above
the SDW transition.
\\
\indent If we assume that $\nu_{Q}$ is roughly inversely
proportional to the unit-cell volume associated with the
structural change which is at most ~$5\%$, the change in
$\nu_{Q}$ gives a spectral broadening of at most $0.02~$T at half
the maximum intensity at $6.6~$T \cite{Hideto-JPSJ-2008}. It
should be borne in mind that in this system the structural change
does not accompany any unit-cell volume change, and therefore the
structural change itself cannot be responsible for such a large
line-broadening at $130~$K. The change in the $\nu_{Q}$ value
from $150~$K to $130~$K is only around $1.43~$MHz, which is
rather unlikely to produce a change of the line width (in terms
of field) by more than an order of magnitude. The changes in the
internal field and the $\nu_{Q}$ and $\eta$ values at $130~$K
indicate that the SDW-type Fe AFM ordering and the structural
transition from TT phase to OT phase occur near-simultaneously.
On the other hand, for the parent compound CeFeAsO the available
results suggest that the structural and AFM Fe ordering take
place at two distinctly different temperatures 150 K and 145 K,
respectively.
\\
\\
\indent As already discussed, the spectral width is an indication of the distribution of the internal field
($\Delta H_{int}$) at the P or As site.
Fig.~\ref{fig:Fit-Field-As_95P_05} (main panel) shows the
variation of the internal field across the magnetic phase
transition. A negligibly small internal field persist from
$220~$K down to $132~$K, but increases abruptly upon cooling
through $130~$K. It is evident that the internal field is nearly
fully developed as soon as ordering sets in and very little
growth in the internal field takes place upon cooling from
$130~$K to $5~$K. The discontinuous tendency of the magnetic
internal field at $130~$K indicates that the SDW transition is
likely to be first-order.
\begin{figure}
\includegraphics[scale=0.9]{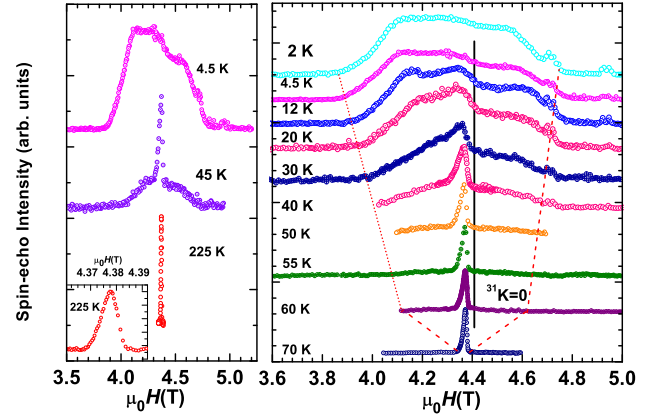}
\caption{\label{fig:PAs_70P_30} $^{31}$P field sweep NMR spectra
at different temperatures for $x=0.3$. The solid vertical line indicates
the position of the Larmor field, and the evolution of the SDW ordering is indicated by the dotted line.}
\end{figure}
\begin{figure}
\includegraphics[scale=0.85]{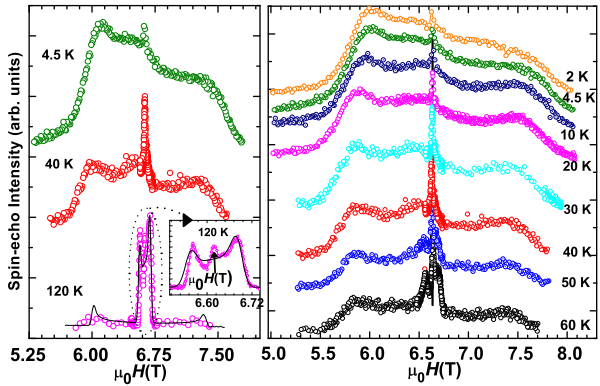}
\caption{\label{fig:AsAs_70P_30} $^{75}$As field sweep NMR
spectra at different temperatures for $x=0.3$. Vertical line
(inset) indicates the position of the Larmor field.}
\end{figure}
\begin{figure}
\includegraphics[scale=0.8]{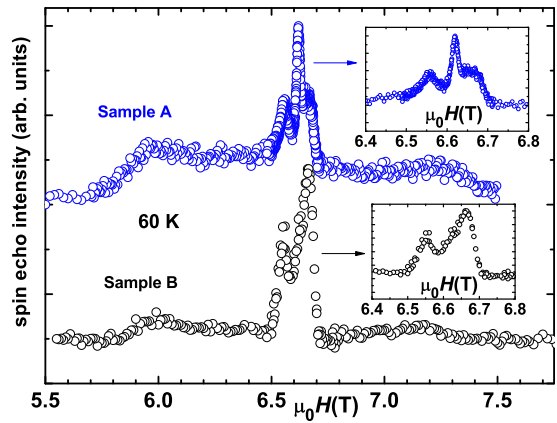}
\caption{\label{fig:Ascom60K} $^{75}$As field sweep NMR spectra at
60 K for $x$=0.3 for two different batches. Inset shows the
quadrupolar splitted central transition for two batches.}
\end{figure}
\subsection{\label{sec:level2}$^{31}$P and $^{75}$As NMR for \textit{x}=0.3: the critical region}
A typical set of $^{31}$P field sweep NMR spectra is shown in
Fig.~\ref{fig:PAs_70P_30} at different temperatures. (Left panel)
Fig.~\ref{fig:PAs_70P_30} shows the $^{31}$P field sweep NMR
spectra at three different temperature. Right-hand panel shows the
temperature variation of $^{31}$P field sweep NMR spectra in the
temperature range $70-2~$K. The NMR spectra change considerably
over this temperature range. The vertical line indicates the
position of Larmor field. At high temperatures (inset in
left-hand panel) a narrow, single anisotropic line is observed as
expected for a TT structure. A moderate anisotropy develops
together with line broadening as the temperature is lowered, and
at the same time there is an overall shift of the resonance
towards lower fields.
\\
\indent The overall line shape is not changed remarkably down to
the temperature 70 K. However, with further lowering of
temperature at around 60 K, all of a sudden a structure develops.
This is more evident below 50 K. The line consists of a
superposition of a smeared anisotropic broadened background and a
narrow single line (At 4.35T). Moreover, the smeared anisotropic
broadened background is gaining intensity and anisotropy upon
lowering the temperature down to 2 K, whereas the intensity of
the single narrow line (At 4.35T) is fading out. And it is hardly
traceable below 20 K.
The small signal observed at around 4.75 T and 4.95 T,  are due
to the presence of small amount Sn impurity, corresponding to the
NMR signals of $^{119}$Sn and $^{117}$Sn isotopes, respectively.
As the sample is prepared by Sn-flux method, therefore it is
likely to be the presence of small amount of impurity in the
sample.
\\
\indent The change of line shape in the $^{31}$P NMR spectra at
around $60~$K is evidently due to a modified SDW- type of Fe AFM
ordering, although the effects of the structural transition from
the TT phase to OT phase cannot be excluded. However this is
unlikely as because synchrotron XRD measurements do not suggest
any structural transition.  The effective width of the broadened
line is about $0.6~$T, which might be used a first estimation of
the internal field of the Fe ordered moment sensed by the P
nuclei. The red dotted lines in Fig.~\ref{fig:PAs_70P_30} are
guides to the eye and track the evolution of the internal field
towards low temperatures. At $60~$K and below, the presence of a
narrow central peak along with the smeared broadened background
suggest co-existence of the paramagnetic and SDW type AFM ordered
phases.
\\
\indent Figure~\ref{fig:AsAs_70P_30} shows the $^{75}$As field
sweep NMR spectra for the $x=0.3$ compound at different
temperatures. In the inset of the left panel, the central part of
the $^{75}$As spectra at $120~$K is shown. The right-hand panel
shows a series of $^{75}$As spectra in the low-temperature range,
and a number of interesting features are noted. At high
temperature the spectra reveal typical second-order quadrupole
splitting. For a single-phase As position, we would expect a
second order split central line as observed for the system
$x=0.05$. However, for $x=0.3$ shown in
Fig,~\ref{fig:AsAs_70P_30}, a relatively sharp narrow line along
with the second-order split central line, at around $6.625~$T, is
also observed. This is indicated by the black arrow in the inset
of Fig.~\ref{fig:AsAs_70P_30} (left-hand panel). This resonance
may conceivably originate from a tiny amount of As-rich impurity
phase. To investigate this conjecture, we have compared these
results (labeled sample A) with a second batch of sample material
(sample B) believed to be of superior analytical quality.
Figure~\ref{fig:Ascom60K} shows an overlay of the $^{75}$As Field
sweep NMR spectra taken at $60~$K for two samples. Inset of
Fig.~\ref{fig:Ascom60K} shows the central transition on a
magnified scale. It is seen that the sharp narrow line at
$6.625~$T is absent in spectrum of sample B, and confirms the
presence of As impurity content in sample A. Moreover, an
impurity of this nature may itself contribute finite line
broadening in the $^{75}$As NMR spectrum. Hence, we refrain from
extracting further detailed information from
Fig.~\ref{fig:AsAs_70P_30}, aside from noting that the
quadrupolar split central transition suffers loss of intensity
towards low temperatures and eventually disappears below $20~$K.
\\
\subsection{\label{sec:level2}$^{31}$P and $^{75}$As NMR for \textit{x}=0.9: the Kondo region}
\begin{figure}
\includegraphics[scale=0.80]{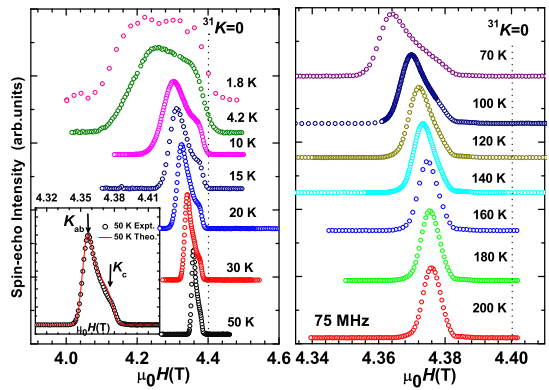}
\caption{\label{fig:SpectraPAs_10P_90} $^{31}$P field sweep NMR
spectra for $x=0.9$ at different temperatures. The vertical dashed
line indicates the Larmor field obtained from the reference
compound. The inset shows a representative powder spectrum, together
with the simulation at $T = 50~$K (arrows indicate the resonance
fields corresponding to the shift values of $^{31}$K$_{ab}$ and
$^{ 31}$K$_{c}$ for the field direction $H\bot c$ and $H\| c$, respectively.}
\end{figure}

\begin{figure}
\includegraphics[scale=1.0]{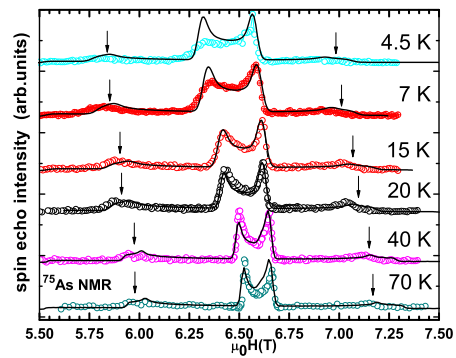}
\caption{\label{fig:SpectraAsAs_10P_90} $^{75}$As field sweep NMR
spectra for $x=0.9$ at different temperatures along with a simulation (see text).
Arrows indicate the position of the satellite
transitions.}
\end{figure}

\begin{figure}
\includegraphics[scale=1.0]{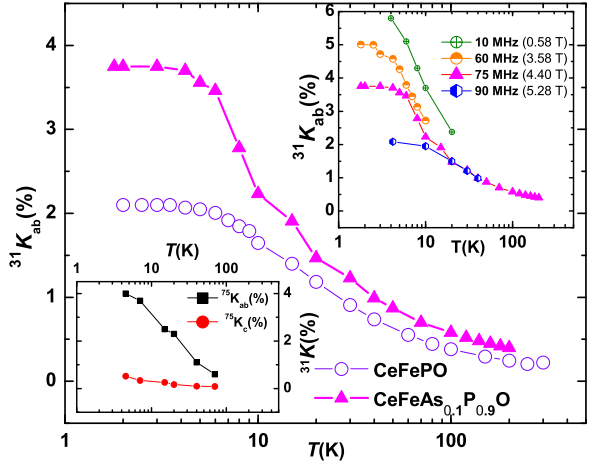}
\caption{\label{fig:ShiftPAs_10P_90} Shift component
$^{31}$K$_{\mathrm{ab}}$ as a function of temperature for CeFePO at $76~$MHz
(data taken from \cite{E. M. Bruning-prl-2008})
and for the CeFeAs$_{0.1}$P$_{0.9}$O at $75~$MHz. Upper inset shows
$^{31}$K$_{\mathrm{ab}}$ at different frequencies (fields). Lower
inset shows $^{75}$K$_\mathrm{{ab}}$ and $^{75}$K$_{\mathrm{c}}$
as a function of temperature at $48~$MHz.}
\end{figure}

A typical set of $^{31}$P field sweep NMR spectra at different
temperatures is shown in Fig.~\ref{fig:SpectraPAs_10P_90}. One
single narrow $^{31}$P-NMR line as expected from the crystal
structure was found at $200~$K (right-hand panel of
Fig.~\ref{fig:SpectraPAs_10P_90}). The line develops strong
asymmetry and increased line-width towards lower temperature
(left-hand panel of Fig.~\ref{fig:SpectraPAs_10P_90}) at
$75~$MHz. Furthermore and in contrast to the other samples in
this study the line shifts strongly with temperature. With
increasing the frequency, onset of line broadening commences
already at high temperatures. The shape is characteristic of a
powder pattern from a spin $I=1/2$ nucleus in a TT symmetry. The
$^{31}$P spectra could be simulated consistently at all
temperatures with shift-tensor components $K_{ab}(T)$ and
$K_{c}(T)$ corresponding to the $H\bot c$ and $H\| c$ directions,
respectively (inset of the Fig.~\ref{fig:SpectraPAs_10P_90}).
Similar to CeFePO $^{31}$$K_{ab}$ shows a strong temperature
dependence whereas $K_{c}$ is almost temperature independent.
Figure~\ref{fig:ShiftPAs_10P_90} shows the variation of
$^{31}$$K_{ab}(T)$ with temperature for CeFePO \cite{E. M.
Bruning-prl-2008}and CeFeAs$_{0.1}$P$_{0.9}$O. Above $30~$K
$^{31}$$K_{ab}$ for the $x=0.9$ sample resembles the
$^{31}$$K_{ab}$ for the $x=1$ sample. It shows CW-like 4f$^{1}$
Ce magnetism. Here, a larger shift value indicates a large
hyperfine field at the P site. However, below $30~$K
$^{31}$$K_{ab}$ for the $x=0.9$ sample deviates significantly
towards lower temperature. The inset shows the field dependence
of $^{31}K_{ab}(T)$. Here, below $10~$K a strong field dependence
of the residual shift for $T \rightarrow$ 0 is observed.
$^{31}K_{ab}(T)$ is decreasing with increasing frequency (field).
This indicates a FM ordered ground state. For a FM system, the
magnitude of the susceptibility should decrease with increasing
the field due to progressive saturation of the magnetization. As
shift is following the susceptibility, the specific field
dependence could be the indication that the system is FM ordered.
Furthermore, line broadening could be the indication of onset of
electronic correlations. This broadening towards lower
temperature could also be associated with the Kondo effect.
\\
\indent Figure~\ref{fig:SpectraAsAs_10P_90} shows the $^{75}$As
field sweep NMR spectra at different temperatures. The obtained
spectra are typical powder patterns with strong quadrupole
coupling. With decreasing temperature the entire spectra are
shifted towards lower fields. Moreover, the spectra become more
anisotropic at lower temperature similar to the case of $^{31}$P
spectra.
All of the $^{75}$As spectra fit consistently in the
whole temperature range and enable estimation of $^{75}$$K_{ab}$
and $^{75}$$K_{c}$. The arrows indicate the satellite
transitions. The obtained fit parameter is $\nu_{Q}=9.27 \pm
0.20$ MHz. Lower inset of Fig.~\ref{fig:ShiftPAs_10P_90} shows
$^{75}$$K(T)$ as a function of temperature. Similar to the
$^{31}$$K(T)$, highly anisotopic behaviour in $^{75}$$K(T)$ is
observed. Here $^{75}$$K_{ab}$ is increased upon lowering the
temperature, whereas $^{75}$$K_{c}$ remains almost independent of
temperature. Similar values of $^{75}$$K_{ab}$ and
$^{31}$$K_{ab}$ indicate that the hyperfine field is similar on
the As and P sites.
\begin{figure}
\includegraphics[scale=.90]{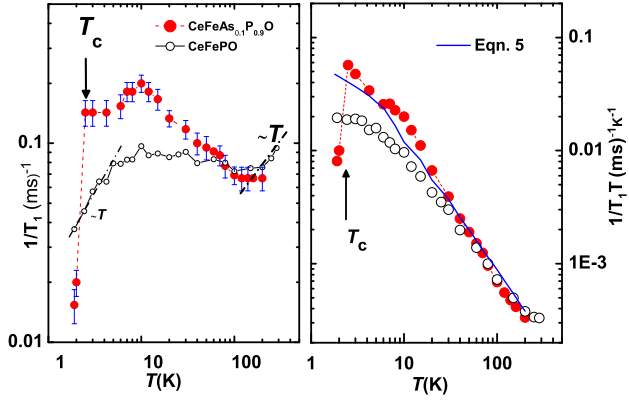}
\caption{\label{fig:T1PAs_10P_90} $^{31}$($1/T_{1}$) as a function
of temperature for CeFePO (data taken from \cite{E. M.
Bruning-prl-2008})and CeFeAs$_{0.1}$P$_{0.9}$O(left).
$^{31}$($1/TT_{1}$) together with the calculated values (solid
line) plotted as function of temperature (right-hand panel).}
\end{figure}
\\
\indent In this subsection we present $^{31}$P spin-lattice
relaxation ($^{31}$(1/$T_{1}$)) data on CeFeAs$_{0.1}$P$_{0.9}$O.
$T_{1}$ measurements were performed at different temperatures at a
frequency of $75~$MHz at the $H\bot c$ position of the
anisotropic NMR line (left arrow in the inset of
Fig.~\ref{fig:SpectraPAs_10P_90}). $^{31}$(1/$T_{1}$) was
obtained by fitting the nuclear magnetization ($m(t)$) recovery
with a standard single exponential function, for a $I=1/2$
nucleus:
\begin{equation}
m(0)-m(t)=m(0)\exp(-t/T_{1})
\end{equation}
Figure~\ref{fig:T1PAs_10P_90} shows the temperature dependence of
$^{31}$(1/$T_{1}$) for the $x=0.9$ sample together with the pure
system CeFePO\cite{E. M. Bruning-prl-2008}. At high temperatures
$^{31}$($1/T_{1})$ of both samples are found to merge whereas
towards lower temperatures a strong increase with decreasing
temperature was found. At around $10~$K, $^{31}$(1/$T_{1}$)
decreases to form a broad maximum. With further cooling at $2~$K
$^{31}$(1/$T_{1}$) is sharply decreased by an order of magnitude.
The sharp decrease of $^{31}$(1/$T_{1}$) indicates the system
undergoes a FM transition at about $2~$K, consistent with bulk
measurements.
\\
\indent The comparison of $^{31}$(1/$T_{1}$) with the parent
compound CeFePO is noteworthy. In contrast to CeFePO, in
CeFeAs$_{0.1}$P$_{0.9}$O the plateau-like region in
$^{31}$(1/$T_{1}$) is absent and long-range FM order is observed.
Crossover from a strong Kondo type of interaction to AFM order
has been studied by NMR as a local probe in a number of cases.
Good examples are systems such as CeCu$_{2}$(Si/Ge)$_{2}$
\cite{{Buttgen-JPSJ-1996},{Buttgen1-physicaB-1997}} and
Ce(Ru/Rh)$_{2}$Si$_{2}$\cite{{Kohra-physicaB-2000},{Ueda-physicaB-1999}}.
On the other hand, systems where doping induces a crossover from
a Kondo type of interaction to long-range FM order are rare. A
distribution of Kondo temperatures or Kondo disorder is likely to
exist in proximity to FM order. One example is
CeRu$_{2}$(Si,Ge)$_{2}$ where the stoichiometric Si compound is a
heavy fermion and the Ge end compound a $T_{c}= 7.5~$K
ferromagnet. For CeRu$_{2}$(Si$_{1-x}$Ge$_{x}$)$_{2}$ the phase
diagram is rather complex and a complex AFM type of phase was
found between the Kondo and the FM region (0.06$\leq x \leq$
0.65)\cite{Haen-physicaB-1999}. Another example is the $^{29}$Si
NMR study in CeSi$_{x}$ where FM order shows up below $x=1.83$
whereas Kondo interaction typifies the behaviour above $x=1.82$
($\leq 2$). Here $^{29}$(1/T$_{1}$) also shows a strong increase
in the vicinity of the FM order\cite{KOHORI-JMMM-1986}. For
CeFeAs$_{0.1}$P$_{0.9}$O we do not have evidence for an AFM type
of ordering, judging by field-dependent NMR investigations. Here,
the Kondo interaction probably competes with FM clusters on a
short time scale. The competition of Kondo and FM type
interactions therefore could be responsible for the non-Korringa
like behavior of CeFeAs$_{0.1}$P$_{0.9}$O.
\\
\indent Furthermore, it should be mentioned that due to the strong
field dependence of the ferromagnetism a strong field dependence
of $^{31}$(1/T$_{1}$) could be expected. Therefore a detailed
study of (1/T$_{1}$) as a function of field is required to
address this problem quantitatively. This topic is not the main
concern of this paper. We would like to add that in a $\mu SR$
study the magnetic volume fraction was found to start increasing
from $10~$K, i.e.\ well above T$_{c}$, which is consistent with
our NMR findings \cite{Johanes-CeFeAsPO-MuSR-2009}. Br\"{u}ning
\emph et al.\ compared NMR results of CeFePO with those of LaRuPO,
LaFeAsO$_{1-x}$F$_{x}$ and CeCu$_{2}$Si$_{2}$ to show that the
$1/T_{1}$ is dominated by the 4f-electron contribution \cite{{E.
M.
Bruning-prl-2008},{{Nakai-JPSJ-2009},{Grafe-PRL-2008},{Aarts-Physica-1983}}}.
In their analysis, they mentioned that a similar temperature
dependence of the relaxation rate were found for the compounds
CeFePO and CeCu$_{2}$Si$_{2}$ and concluded a similar spin
fluctuation relaxation mechanism in both compounds. Furthermore,
the strong correlation effects in CeFePO originate from the Ce-4f
electrons rather than the from the Fe-3d electrons.
\\
\indent From the fluctuation dissipation theorem for a localized
4f-electron moment system the $1/T_{1}$ can be written as
\cite{Pennigton-NMR-1990}
\begin{equation}
1/T_{1}T\propto \gamma^{2}\sum_q
A^2_{hf}(\mathbf{q},w)\frac{\chi''(\mathbf{q},w)}{w}
\end{equation}
where A$_{hf}$ is the effective hyperfine coupling and
$\chi''(\mathbf{q},w)$ is the absorptive part of dynamical spin
susceptibility.
\\
\indent In a simplified picture with a small number of Ce
neighbouring atoms, a situation which prevails in the present
system (P/As is coupled to 4 Ce neighbors), one may neglect the
$\mathbf{q}$ dependent contribution. Under these circumstances,
1/($T_{1}T$) in a Kondo lattice should be dominated by the
contribution of the 4f electrons, the latter which may be
approximated as proportional to the ratio of the static
susceptibility $\chi(T)$ and the effective relaxation rate
$\Gamma(T)$ of the 4f electrons,
\begin{equation}
(1/T_{1}T)_{4f}\propto\chi(T)/\Gamma(T))
\end{equation}
\cite{{Nakamura-JPSJ-sup-1996},{Buttgen-JPSJ-1996},{Kuramoto-OXFORD-2000},{MacLaughlin-hyperfine-1996}}.
Further on for Kondo systems $\Gamma(T) \propto \sqrt{T}$ is
valid for $T>T_{K}$ \cite{D. L. Cox-NMR-J-app. Phys.-1985}, where
as for $T\ll T_{K}$ a constant value  $\Gamma_{0}\sim K_{B}T_{K}$
was found. We calculated the temperature dependence of
\begin{equation}
(1/T_{1}T)_{4f}=C^{31}K_{ab}(T)/\sqrt{T} \label{eun5}
\end{equation}
where $C$ is a temperature-independent parameter. We added a
small temperature-independent contribution $(1/T_{1}T)_{CE}=0.0001
(ms)^{-1}K^{-1}$ to account for the contribution of the
conduction electrons at high temperature where the contribution
of the 4f electrons becomes negligible. The solid line in
Fig.~\ref{fig:T1PAs_10P_90} (right-hand) illustrates our
calculation. The experimental data of CeFeAs$_{0.1}$P$_{0.9}$O are
adequately described by this approach. It seems obvious that the
onset of Kondo interactions in this doped compound is similar to
that in pure CeFePO. Additionally Fig.~\ref{fig:T1PAs_10P_90}
shows that in the temperature range $40-2~$K the magnitude of
$^{31}(1/T_{1}T)$ is higher than what is the case in CeFePO,
which is consistent with the magnetic susceptibility and Knight
shift data. Furthermore it should be mentioned that the
$^{31}($1/T$_{1}$) power law (eqn.\ref{eun5}) observed here
differs significantly from what was found for itinerant
ferromagnets like ZrZn$_{2}$, where 1/T$_{1}T\sim \chi (T)$ is
valid. This indicates that the system is not a typical
Moriya-type ferromagnet.
\\
\indent In contrast to CeFePO, an additional long-range FM
ordering of Ce takes place in the present As-doped system. The
residual shift exhibits a strong field dependence which is
characteristic of FM ordering at low temperature. This is
consistent with $C_{p}(T)$ measurements where ordering at
$T_{c}=2~$K was found \cite{Jesche-PHD-Thesis-2011}. The line
broadening in NMR data could be due to reduction of $T_{2}$ due
to Kondo and/or FM interactions. To conclude, the P-rich $x=0.9$
sample could be identified as a heavy Fermion system with a FM
ground state. This presents a rather unique situation among
correlated electron systems.
\subsection{\label{sec:level2}Comparison and discussion}
After presenting the detailed NMR results for three P
concentrations of CeFeAs$_{1-x}$P$_{x}$O, we now compare and
discuss our findings. Table \ref{tab:table1} collects the
important results obtained from the NMR study for three P
concentrations. For $x$=0.05 : AFM (Fe)+ structural transition at
130 K, with higher P content of $x$=0.30 : AFM (Fe) ordering
suppressed to 70 K and no structural transition, $x$=0.90 : no AFM
Fe ordering, no structural transition, FM ordering of Ce. In a
recent study of Clarina de la Cruz \emph et. al.\ it is claimed
it is not possible to separate out the structural from the
magnetic phase transition for $x\geq$ 0.05 within experimental
resolution \cite{Clarina de la Cruz-condmat-CeFeAs1-xPxO-2009}.
The alloy system CeFeAs$_{1-x}$P$_{x}$O becomes paramagnetic in
the TT structure, like CeFePO, above $x\approx 0.4$. However,
bulk measurements suggest that above $x\approx 0.4$, Ce order FM.
Furthermore according to Clarina de la cruz \emph et. al.\, for
$x=0.05$ two transitions (magnetic and structural) take place at
around $140~$K, whereas our NMR investigations describe a similar
occurrence albeit at $T=130$ K, -a result which is in line with
the resistivity data shows the anomaly close to $T$=130~K
\cite{Jesche-PHD-Thesis-2011}. The occurrence of both transitions
(magnetic and structural) at the same temperature for the present
system  is in contrast to all other doping series where with
increasing doping at the Fe/As site structural and magnetic
transition is getting separated. Therefore present NMR results
apart from confirming the magnetic and structural transitions,
additionally provide insightful information. Here it should be
noted that NMR results describe the drastic change of internal
field due to Fe $3d$ ordered moment at $T$=130~K. Basically, this
change of internal field is the consequence of a changing
magnetic order parameter. Our data are supportive of the Fe-AFM
SDW type transition to be first-order.
\\
\indent According to neutron scattering results for $x=0.3$, the
structural transition and Fe AFM ordering ($T_{\mathrm{N}}$) take
place at $\approx$85~K, which is relatively higher than the
suggested result from bulk measurements $\approx$70~K. On the
other hand, NMR investigations, in line with $\mu SR$ results
\cite{Johanes-CeFeAsPO-MuSR-2009}
 described this phase transition consistent with the findings of bulk measurements.
 From the $^{31}$P NMR study there is evidence for a phase
separation in terms of paramagnetic and AFM-SDW type phases. This
rules out the possibility of termination to a quantum critical
point with increasing P concentration at the $x=0.3$ region.
\\
\indent In the following paragraph, we present a comparison of the
relative change in internal field upon P doping (0.05
$\rightarrow$ 0.3) as estimated from the $^{31}$P and $^{75}$As
NMR study. The field ratios are given by
\begin{eqnarray}
\frac{(^{31}\Delta H_{\mathrm{int}})_{0.05}}{(^{31}\Delta
H_{int})_{0.3}}= \frac{1.132  \mathrm{T}}{0.6 \mathrm{T}}\simeq
1.9,
\\
\frac{(^{75}\Delta H_{\mathrm{int}})_{0.05}}{(^{75}\Delta
H_{\mathrm{int}})_{0.3}}= \frac{1.9 \mathrm{T}}{1.1
\mathrm{T}}\simeq 1.727 \label{eq:relative75As}~~.
\end{eqnarray}
On the other hand, a relative change of the Fe static ordered moment,
moving from $x=0.05$ to $0.3$, has been reported from neutron
scattering and is as follows,
\begin{equation}
\frac{(\mu_{Fe})_{0.05}}{(\mu_{Fe})_{0.3}}\Longrightarrow
\frac{0.8 \mu_{B}}{0.4 \mu_{B}}\simeq 2~~.
\end{equation}
Therefore the relative change of Fe static ordered moment for the two
P concentration values ($x=0.05, 0.3$) and the relative change of the
estimated internal field for the same P concentration are in close agreement.
\\
\indent The magnitude of the estimated internal field (in our
approach) from powder $^{31}$P and $^{75}$As spectra, in
principle, may vary because for $^{75}$As spectra, below the
Fe-ordering, satellites may introduce additional line broadening.
This makes the $^{75}$As NMR spectra even more complicated.
Therefore it is not possible to isolate the effect of internal
field to the central transition. As a result, the estimated
internal field from the $^{31}$P and $^{75}$As NMR spectra may
vary for a specific P concentration. Nonetheless, the relative
change of internal field for two different P concentrations
should be independent of the probed nuclei, which is indeed the
case here.
\\
\indent For the $x=0.05$ and $0.3$ compositions an additional line
broadening accompanies the Ce magnetism at low temperatures. Well
below the Fe AFM transition the Fe ordered moment should
saturate. Thus, this cannot produce any additional line
broadening in the spectra. Therefore the additional broadening in
$^{75}$As spectra originates from the Ce magnetism for $x=0.05$. A
similar situation prevails in the case of the $x=0.3$ sample.
\\
\indent By comparison, the magnetic transition at Fe-AFM ordering
is rather sharp in the $x=0.05$ case, compared to $x=0.3$. The
line shape changes rather drastically for the $x=0.05$ sample,
while a gradual evolution of line shape is observed in $x=0.3$.
The magnetic transition in the $x=0.05$ system is likely to be
first-order. For $x=0.3$ the internal field does not affect the
line shape as strongly compared to the $x=0.05$ system. By
contrast, it suddenly develops a distinct structure at full width
of quarter maximum position at around $70~$K, and the intensity
develops with lowering temperature in a manner which suggests
either that the ordered moment of Fe is still not fully saturated
and/or that not all of the Fe moments participate in ordering
just below $70~$K. This leads to the conjecture that paramagnetic
and ordered phases may co-exist in a presumably inhomogeneous
distribution. The magnetic volume fraction of this compound just
below the ordering temperature may be somewhat less than $1$.
Nonetheless, at sufficiently low temperatures a magnetic volume
fraction of $1$ is eventually achieved.
\\
\indent For $x=0.9$ the TT phase persist throughout the entire
temperature range. In contrast to the $x=1$ sample a long range FM
order is likely to be confirmed by $^{31}$K measurements.
Additionally, the $^{31}$(1/T$_{1}$) results indicate a complex
interplay of FM and Kondo type of fluctuations in the proximity
of the long range FM order. To study the very interesting
cross-over from Kondo to FM type of interaction more detail
studies are required.
\\
\indent Next, we compare the three concentrations ($x=0.05$,
$0.3$, and $0.9$) in the context of shift results. The magnitude
of the shift was found to increase with increasing P
concentration. For $x=0.05$ a near temperature-independent shift
is observed. For $x=0.3$ the line position is shifted slightly
with lowering the temperature indicating a small shift value. A
cancellation of hyperfine fields at the P/As site may be
responsible for a weakly temperature dependent shift. This is, in
fact, very likely because the conduction electron polarization
from the $4f^{1}$ ion produces a positive field whereas the core
polarization from the $3d$ ion usually results in a negative
hyperfine field. However, with increasing P concentration the
effect of Fe $3d$ moments is reduced. This enhances the shift.
Therefore it is clear that in the As rich side of the
CeFeAs$_{1-x}$P$_{x}$O alloy the contributions of Ce-$4f$ and
Fe-$3d$ magnetism are significant, whereas in the P rich side the
Ce $4f$ magnetism dominates. Cancellation could also be possible
to come by between the moments of the Fe-sublattice themselves
with P/As either occupying symmetric positions in the magnetic
lattice or random positions in an incommensurate SDW.
\\
\indent Comparing the $^{31}$P NMR spectra for three different
concentration at low temperature, a noticeable difference in the
effective line width is observed close to $5~$K. With increasing
P concentration the line width decreases. This is consistent with
the fact that contribution of Fe magnetism is lowered with
increasing P concentration.
\begin{table*}
\caption{\label{tab:table1} Comparison of results obtained from
different P concentration.}
\begin{ruledtabular}
\begin{tabular}{ccccc}
 & &$x$=0.05&$x$=0.3&$x$=0.9\\ \hline
 T$_{N}(Fe)$ && 130 K & 70 K & No Fe ordering \\
 T$_{N,C}$(Ce) && Additional line broadening& Additional line broadening& 2 K (FM)\\
 Structural transition &&130 K (TT$\rightarrow$OT)\footnote{TT-Tetragonal Phase, OT-Orthorhombic
 Phase.}& No  transition & No Transition(TT)\\
\end{tabular}
\end{ruledtabular}
\end{table*}

\section{Summary and conclusion}

To conclude, we have prepared polycrystalline samples of
CeFeAs$_{1-x}$P$_{x}$O by a Sn-flux technique. A systematic study
of $^{31}$P and $^{75}$As NMR was conducted on the $x=0.05$,
$0.3$ and $0.9$ members of this series. 1).\ For the
CeFeAs$_{0.95}$P$_{0.05}$O compound a drastic change of the line
width at $130~$K indicates AFM ordering of Fe and the structural
change from TT to OT. Associating linewidth with the internal
field of the system, a large change of the internal field
evidences AFM (SDW type) transition which is likely to be
first-order. Small and nearly constant shift values are found in
$^{31}$P and $^{75}$As NMR and ascribed to competing mechanism of
the $4f$ and $3d$ magnetism of Ce and Fe respectively.
Simulations of powder spectra are complex below the SDW
transition because lineshapes are influenced by coinciding
SDW-magnetic and structural phase transitions. 2).\ On the other
hand, for the CeFeAs$_{0.7}$P$_{0.3}$O compound the evolution of
the Fe-SDW type order close to $70~$K corroborates the results of
bulk measurement and $\mu SR$. The complicated line shapes in
spectra of this system do not permit unambiguous fitting to be
performed. The line shape is, nonetheless, in evidence of a phase
separation (paramagnetic and ordered phase) taking place. A
considerable anisotropy develops upon cooling. 3).\ In contrast
to CeFePO, in CeFeAs$_{0.1}$P$_{0.9}$O additional magnetic
ordering develops. Field-dependent shift results give the evidence
of FM ordering. Above the ordering $^{31}$(1/T$_{1}$) shows
unconventional, non-Korringa like behaviour which indicates a
complex interplay of Kondo and FM fluctuations.
\\
\indent The present system contributes valuable insights about
incipient magnetic order in the presently intensively studied
iron pnictide family of compounds, but warrants thorough
investigations in its own right due to the unusual emergence of
cooperative ferromagnetism within a Kondo lattice of
local Ce moments that are generically coupled antiferromagnetically to the conduction electrons.\\

\begin{acknowledgments}
We are thankful to Dr. C. Krellner and Prof. Q. Si for stimulating
discussion at the very beginning of this project. We are grateful
to Prof. A. Strydom for carefully reading and considerably
improving this manuscript.
\end{acknowledgments}
\bibliography{CeFeAsO}

\begin{thebibliography}{10}
\bibitem{Kamihara-2008}
Kamihara Y, Watanabe T,  Hirano M, and  Hosono H 2008 
\textit{J. Am. Chem. Soc.} \textbf{130}, 3296
\bibitem{Chen-2008}
Chen X H, Wu T, Wu G, Liu R H, Chen H and Fang D F 2008 
\textit{Nature} \textbf{453}, 761
\bibitem{Jun Zhao-nature materials-2008}
 Zhao Jun,  Huang Q, Clarina de la Cruz,  Li S,  Lynn J W, Chen Y,
 Green M A, Chen G F,  Li G,  Li Z,  Luo J L,  Wang N L and  Dai P 2008
\textit{Nature Materials} \textbf{7}, 953-959.
\bibitem{Z. A. Ren-2008}
 Ren Z A,  Yang J, Lu  W,  Yi W,  Che G -C,  Dong X-L, 
Sun L-L, and  Zhao Z X 2008 \textit{Mater. Res. Innovations} \textbf{12}, 105 
\bibitem{Chen-Li-2008}
 Chen G F, Li  Z,  Wu D,   Li G,  Hu W Z,  Dong J,  Zheng P,  Luo J L and 
Wang N L 2008  \textit{Phys. Rev. Lett.} \textbf{100}, 247002 
\bibitem{J. Yang-Supercond.Sci. Technol.-2008}
Yang J, Li Z C, Lu W, Yi W, Shen X L, Ren Z A, 
Che G C, Dong X L, Sun L L, Zhou F and  Zhao Z X 2008 \textit{Supercond.Sci. Technol.} \textbf{21}, 082001 
\bibitem{J. G. Bos- Chem. Commun.-2008}
 Bos J G,  Penny G B S,  Rodgers J A,  Sokolov D A, 
Huxley A D and  Attfield J P 2008 \textit{Chem. Commun. (Cambridge)}, 3634 
\bibitem{Krellner-2007}
 Krellner C,  Kini N S,  Br\"{u}ning E M,  Koch K,  Rosner H,  Nicklas M,
 Baenitz M and  Geibel C 2007 \textit{Phys. Rev. B} \textbf{76}, 104418 
\bibitem{Krellner-Crystal frowth-2007}
 Krellner C and  Geibel C 2007 \textit{J. Cryst. Growth} \textbf{310}, 1875-1880
\bibitem{C. Krellner-prl-2007}
Krellner C,  F\"{o}rster T,  Jeevan H,  Geibel C and  Sichelschmidt J 2008
\textit{Phys. Rev. Lett.} \textbf{100}, 066401 
\bibitem{E. M. Bruning-prl-2008}
 Br\"{u}ning E M,  Krellner C,  Baenitz M, Jesche  A,  Steglich F and  Geibel C 2008
\textit{Phys. Rev. Lett.} \textbf{101}, 117206 
\bibitem{Anton-CeFeAsO-NJP-2009}
Jesche A,  Krellner C,  Souza M de, Lang M and
Geibel C 2009 \textit{New J. Phys.} \textbf{11}, 103050 
\bibitem{Anton-CeFeAsPO-to be published-2009}
A. Jesche. (unpublished)
\bibitem{Chi S-prl-2008}
Chi S, Adroja D T, Guidi T, Bewley R, Li S, Zhao J, Lynn J W, Brown C M, Qiu Y,
Chen G F, Lou J L, Wang N L and Dai P 2008 \textit{Phys. Rev. Lett.} \textbf{101}, 217002 
\bibitem{Maeter condmat-2009}
Maeter H, Luetkens H, Pashkevich Yu G,  Kwadrin A, Khasanov R,  Amato A, Gusev A A,  Lamonova K
V, Chervinskii D A, Klingeler R, Hess C, Behr G, B\"{u}hchner B and  Klauss H H 2009
\textit{Phys. Rev. B} \textbf{80}, 094524 
\bibitem{Shen V. Chong-2009}
Chong S V, Mochiji T, Sato S and  Kadowaki K 2008
\textit{Proc. Int. Symp. Fe-Pnictide Superconductors J. Phys. Soc. Jpn.} \textbf{77} Suppl. C,  27-31 
\bibitem{Takeshitai-JPSJ-2009}
Takeshitai N,  Miyazawa K, Iyo A, Kito H and  Eisaki H 2009 \textit{J. Phys. Soc. Jpn.} \textbf{78}, 065002 
\bibitem{Jianhui Dai-nature mater-2008}
Dai J, Si Q, Zhu  Jian-Xin and  Abrahams E 2009 \textit{Proc Natl. Acad. Sci} \textbf{106}, 4118-4121 
\bibitem{Yongkang Luo-condmat-2009}
Luo Y, Li Y, Jiang S, Dai J, Cao G and  Xu Z  2010 \textit{Phys. Rev. B} \textbf{81}, 134422 
\bibitem{Clarina de la Cruz-condmat-CeFeAs1-xPxO-2009}
Clarina de la Cruz, Hu  W,  Li S,  Huang Q,  Lynn J W,  Green M A,  Chen G F,  Wang N L,  Mook H A,  Si Q and Dai P 2010
\textit{Phys.Rev. Lett.} \textbf{104}, 017204 
\bibitem{Hideto-JPSJ-2008}
Fukuzawa H,   Hirayama K,  Kondo K,  Yamazaki T,
Kohori Y,  Takeshita N,  Miyazawa K,
Kito H, Eisaki H and  Iyo A  2008 \textit{J. Phys. Soc. Jpn.} \textbf{77}, 093706  
\bibitem{Baek-prb-2009}
 Baek S H, Curro N. J, Klimczuk T, Bauer E D, Ronning F and Thompson J D  2009 \textit{Phys. Rev. B} \textbf{79}, 052504
\bibitem{S.-H. Baek-PRL-2009}
Baek S H, Lee H, Brown S E, Curro N J, Bauer E D, Ronning F, Park T and  Thompson J D 2009 \textit{Phys. Rev. Lett.} \textbf{102}, 227601 
\bibitem{Ghoshray-2009}
Ghoshray A, Pahari B, Majumder M, Ghosh M, Ghoshray K, Bandyopadhyay B, Dasgupta P,
Poddar A and Mazumdar C 2009 \textit{Phys. Rev. B} \textbf{79}, 144512 
\bibitem{Mukuda-JPSP-2008}
Mukuda H, Terasaki N, Kinouchi H, Yashima M, Kitaoka Y,
Suzuki S, Miyasaka S, Tajima S, Miyazawa K, Shirage P, 
Kito H, Eisaki H and Iyo A 2008 \textit{J. Phys. Soc. Jpn.} \textbf{77}, 093704
\bibitem{G.C.Carter-Bennett}
Carter G C, Bennett L H and Kahan D J : \textit{Metallic
Shifts in NMR (Pergamon Press, Oxford, 1977)}

\bibitem{Kitagawa-BaFe2As2-NMR77}
 Kitagawa K,  Katayama N, Ohgushi K, Yoshida M and Takigawa M 2008
\textit{J. Phys. Soc. Jpn.} \textbf{77}, 114709 
\bibitem{Jegli-PRB-2009}
 Jegli\`{e} P,  Bos J W G,  Zorko A,  Brunelli M,  Koch K,  Rosner H,  Margadonna S
 and  Ar\`{e}on D  2009 \textit{Phys. Rev. B} \textbf{79}, 094515 
\bibitem{George H. Stauss-JCP-1964}
 Stauss George H 1964
\textit{J. Chem. Phys.} \textbf{40}, 1988 
\bibitem{Kenichiro Tatsumi-JPSJ78-2009}
 Tatsumi K,  Fujiwara N,  Okada H,  Takahashi H,  Kamihara Y,  Hirano M and  Hosono H 2009
\textit{J. Phys. Soc. Jpn.} \textbf{78}, 023709 
\bibitem{Nakai-JPSJ-2009}
 Nakai Y,  Ishida K,  Kamihara Y,  Hirano M and
 Hosono H 2008 \textit{J. Phys. Soc. Jpn.} \textbf{77}, 073701
\bibitem{Grafe-PRL-2008}
 Grafe H J, Paar D,  Lang G,  Curro N J,  Behr G,  Werner J,  Hamann-Borrero J,  Hess C,  Leps N,
 Klingeler R and  B\"{u}chner B  2008 \textit{Phys. Rev. Lett.} \textbf{101}, 047003 
\bibitem{Aarts-Physica-1983}
 Aarts J,  deBoer F and  MacLaughlin D E  1983
\textit{Phys. Rev. B}  \textbf{121 ($B+C$)} (Amsterdam), 162
\bibitem{Pennigton-NMR-1990}
Pennigton C H and  Slichter C P 1990 \textit{in Physical Properties of High
Temperature Superconductors II, edited by D.M. Ginsberg}
~World Scientific, Singapore
\bibitem{Nakamura-JPSJ-sup-1996}
 Nakamura H, Shiga M, Kitaoka Y, Asayama K and 
Yoshimura K 1996 \textit{J. Phys. Soc. Jpn.} \textbf{65}, Suppl. B, 168 
\bibitem{Buttgen-JPSJ-1996}
B\"{u}ttgen N,  B\"{o}hmer R,  Krimmel A and  Loidl A 1996 \textit{Phys.
Rev. B} \textbf{53}, 5557 
\bibitem{Buttgen1-physicaB-1997}
 B\"{u}ttgen N, Krug von Nidda H A and Loidl A 1997
\textit{Physica B}, \textbf{230-232}  590-592
\bibitem{Kohra-physicaB-2000}
Kohara T, Mishina S, Ueda K, Yamamoto Y and  Miyako Y 2000
\textit{Physica B}, \textbf{284-288} 1271-1272
\bibitem{Ueda-physicaB-1999}
Ueda K, Mishina S, Kohara T, Yamamoto Y and Miyako Y 1999
\textit{Physica B}, \textbf{259-261} 83-84
\bibitem{Haen-physicaB-1999}
Haen P, Bioud H and  Fukuhara T 1999
\textit{Physica B} \textbf{259-261} 85-86
\bibitem{Kuramoto-OXFORD-2000}
Kuramoto Y and Kitaoka Y 2000 \newblock {\em Dynamics of Heavy
Electrons} (Oxford Science, New York)
\bibitem{MacLaughlin-hyperfine-1996}
MacLaughlin D E 1989 \textit{Hyperfine Interact.} \textbf{49}, 43 
\bibitem{D. L. Cox-NMR-J-app. Phys.-1985}
 Cox D L, Bickers N E and  Wilkins J W 1985 \textit{J. Appl. Phys.} \textbf{57}, 3166 
\bibitem{Johanes-CeFeAsPO-MuSR-2009}
Spehling  J \textit{et al.} (Unpublished)
\bibitem{Yoshikazu-PRB-2004}
Tabata Y, Taniguchi T, Miyako Y, Tegus O, 
Menovsky A A and   Mydosh J A 2004 \textit{Phys. Rev. B} \textbf{70}, 144415 
\bibitem{KOHORI-JMMM-1986}
 Kohori Y,  Kohra T,  Asayama K,  Satoh N,  Yashima H,  Mori H and
 Satoh T 1986 \textit{J Mag. Magn. Mat.} \textbf{54-57}, 437-438
\bibitem{Jesche-PHD-Thesis-2011}
A. Jesche PhD thesis (Unpublished)

 
\end{thebibliography}

\end{document}